\documentclass{llncs}
\usepackage{graphicx}

\usepackage{amssymb}
\usepackage{amsmath}
\usepackage{colortbl}
\usepackage{setspace}
\usepackage{color}
\usepackage{url}
\usepackage{subfigure}
\usepackage{multirow}
\usepackage[all]{xy}
\usepackage{algorithm}
\usepackage{wrapfig}

\usepackage{algpseudocode}
\usepackage{latexsym}
\usepackage{makeidx}
\usepackage{cite}

\usepackage{mathtools}
\DeclarePairedDelimiter{\ceil}{\lceil}{\rceil}
\DeclarePairedDelimiter{\floor}{\lfloor}{\rfloor}

\newcommand\SECPAR{\kappa}
\newcommand\sol{{\sf PrivLoc}}
\newcommand\MK{\mathit{k}}
\newcommand\LENGTH{\mathit{L}}
\newcommand\COLUMNS{\mathit{m}}
\newcommand\ROWS{\mathit{n}}
\newcommand\newx{\mathit{newx}}
\newcommand\newy{\mathit{newy}}

\begin{document}

\mainmatter
\title{PrivLoc: Preventing Location Tracking in \\Geofencing Services}

\author{Jens Mathias Bohli, Dan Dobre, Ghassan O. Karame, Wenting Li}
\institute{NEC Laboratories Europe, Germany\\\email{firstname.lastname@neclab.eu}}

\maketitle

\begin{abstract}
Location-based services are increasingly used in our daily activities. In current services, users however have to give up their location privacy in order to acquire the service.

The literature features a large number of contributions which aim at enhancing user privacy in location-based services. Most of these contributions obfuscate the locations of users using spatial and/or temporal cloaking in order to provide $k$-anonymity. Although such schemes can indeed strengthen the location privacy of users, they often decrease the service quality and do not necessarily prevent the possible tracking of user movements (i.e., direction, trajectory, velocity). With the rise of Geofencing
applications, tracking of movements becomes more evident since, in these settings, the service provider is not only requesting a single location of the user, but requires the movement vectors of users to determine whether the user
has entered/exited a Geofence of interest.

In this paper, we propose a novel solution, \sol, which enables the privacy-preserving outsourcing of Geofencing and location-based services to the cloud without leaking any meaningful information about the location, trajectory, and velocity of the users. Notably, \sol{} enables an efficient and privacy-preserving
intersection of movement vectors with any polygon of interest, leveraging
functionality from existing Geofencing services or spatial databases. We analyze the security and privacy provisions of \sol{} and we evaluate the performance of our scheme by means of implementation. Our results show that
the performance overhead introduced by \sol{} can be largely tolerated in realistic deployment settings.

\end{abstract}

\section{Introduction}

Location-based services (e.g., Foursquare~\cite{fousquare} and Yelp~\cite{yelp}) are gaining increasing importance recently. Several applications enable users (e.g., using mobile devices) to discover
and communicate their locations to a server in the cloud; in turn, the
server uses this information to return data relevant at the users' locations. For instance, a number of existing services can only be acquired by users who are located within
a specific geographical area; these include banking services, Youtube, and content delivery services, among many others. Location information also proves to be useful for a number of security-critical services
such as police investigations, e-voting, etc.

However, while many devices (e.g., smartphones, tablets) are capable of discovering and reporting their locations, a considerable number of users shy away
from reporting their locations in the fear of being tracked or profiled by service providers~\cite{Barkhuus03location-basedservices}. This problem is even more evident when the service provider wishes to outsource his spatial services to the cloud (e.g.,~\cite{valtus}).
Service providers have considerable incentives to rely on hosted services in the cloud, since this enables them to maximize the availability of the service while minimizing the costs for acquisition of hardware and operation. Indeed, the cloud offers a low barrier for small and medium enterprises to offer location-based services and enables its clients to avoid huge upfront investments to accommodate for peak usage. However, hosting the spatial service in the cloud raises serious privacy concerns with respect to the leakage of client location information to the cloud provider.

The literature comprises a plethora of contributions which strengthen the privacy of users in location-based services. Most of these contributions focus on anonymizing user locations by means of a trusted
 location anonymizer server~\cite{Bamba:2008:SAL:1367497.1367531,Gruteser:2003:AUL:1066116.1189037,Kalnis:2007:PLI:1313047.1313203,Mokbel:2006:NCQ:1182635.1164193,Yiu:2010:ESS:1825238.1825264}. After registering with the service, users can send their exact locations to the server, which ``blurs'' these location reports and sends the cloaked location to a remote database server.
 The server also filters the database's response and subsequently sends the exact answers back to the users when needed. Existing location anonymization techniques can be categorized according to three different approaches: \emph{(i)} inserting false dummies~\cite{Kido05ananonymous}, where
 the server sends $n$ location updates for each location reported by the user ($(n-1)$ reports of which are dummy), \emph{(ii)} location blurring~\cite{Chow:2007:EPC:1784462.1784477,Bamba:2008:SAL:1367497.1367531,Gruteser:2003:AUL:1066116.1189037,Kalnis:2007:PLI:1313047.1313203,DBLP:conf/infocom/WangXHZLX12} where the location of the user is blurred into a
 spatial area (using spatial or temporal cloaking), and \emph{(iii)} landmark obfuscation~\cite{Hong:2004:APU:990064.990087} where the server sends the location of a near-by landmark instead of the location of the user.

 While these techniques can provide users with $k$-anonymity guarantees, existing techniques \emph{(i)} often require changes to the database server in order to efficiently process
 the anonymized queries, or \emph{(ii)} reduce the accuracy of the location-based service (e.g., when relying on spatial cloaking), or \emph{(iii)} do not prevent location tracking~\cite{Mokbel:2006:NCQ:1182635.1164193}. Location tracking
 could be performed e.g., by inferring the direction of the movement/path followed by the user, the velocity of
 the user, etc., which might de-anonymize users. Such information leakage is particularity damaging in location-based services where service providers are interested in the events of users entering/exiting
 a given area (e.g., Geofencing applications~\cite{Sheth:2009:GCW:1560004.1560029,geofencing,geofencing1,Shekhar:1999:SDR:627320.627970,Myllymaki:2003:HSI:775152.775168}). These applications are gaining increasing importance
 for e.g., targeted advertisements, and typically take as inputs
 vectors of movements
 performed by users and enable service
 providers to extract various statistics about their customers, such as visit durations, start, end, etc.

 In this paper, we address this problem, and we propose a novel solution, \sol, which efficiently enables the privacy-preserving use of Geofencing services in the cloud without incurring any modifications to spatial indexing techniques, and without
 leaking any meaningful information about the location, trajectory, and/or velocity of the users to the cloud provider. More specifically, we consider a typical Geofencing setting, whereby
 a spatial database hosted on the cloud receives regular movement vectors from users, and checks if these movements cross a Geofenced area which has been subscribed to within the database. If so, the provider issues a notification
 informing the appropriate subscriber that a user has exited/entered the subscribed area of interest. Note that this can be achieved in existing spatial databases by querying for intersection between the user movement vector and the
 Geofenced area. \sol{} leverages the presence of a private trusted service which performs efficient specially-crafted transformations of location reports to interface with the cloud-hosted Geofencing service without exposing
the privacy of users. As such, \sol{} can be used by companies and individuals to prevent information leakage towards spatial databases hosted in clouds, such as Google, Amazon, etc.
In this respect, \sol{} enables a privacy-preserving intersection of movement vectors with any polygon of interest, while leveraging functionality from existing cloud-based spatial databases. We analyze the security and
 privacy provisions
of \sol{} and we evaluate its performance in a realistic setting. Our results show that our scheme scales well with the number of users and subscriptions in the system and does not incur considerable computational overhead on the trusted server.

The remainder of the paper is organized as follows. In Section~\ref{sec:model}, we outline our system and adversarial model. In Section~\ref{sec:solution}, we introduce and analyze our solution, \sol, which efficiently enables users to
acquire privacy-preserving Geofencing services hosted in the cloud. We evaluate its performance by means of an implementation in Section~\ref{sec:implementation}.
 In Section~\ref{sec:related}, we overview related work in the area, and we conclude the paper in Section~\ref{sec:conclusion}.

\section{Model}\label{sec:model}

In this section, we describe our system and adversarial model, and we outline the security requirements that our solution should satisfy.

\subsection{Spatial Databases}\label{subsec:geo}

Spatial databases are instances of databases optimized to store and query data which represents objects defined in a geometric space.
Examples of spatial databases include MongoDB, MySQL, PostgreSQL, among others.
To efficiently handle and store spatial data, spatial databases rely on a Spatial Database Management System (SDBMS) which extends upon the capabilities of a traditional database management system.
SDBMS typically supports three types of queries: \emph{(i)} set operators (e.g., disjoint, touch, contains), \emph{(ii)} spatial analysis (e.g., distance, intersection), and \emph{(iii)} other basic functions such as envelope, boundary, etc.
This is efficiently achieved through the reliance on spatial indices (e.g., R-tree, X-tree, GiST). For instance, R-trees~\cite{Rtree} represent objects with their minimum bounding rectangle in the next higher level of the tree.
The main intuition here is that a query which does not intersect the bounding rectangle also cannot intersect any of the contained objects.

While there are a number of spatial indexing techniques, all techniques require the database server to check a series of coordinate equalities and inequalities in order to determine
its spatial index. This clearly poses a problem when dealing with encrypted data objects. For instance, standard encryption
of the coordinate system with a semantically secure cryptosystem such as AES, would not preserve any relationship (i.e., to check for equality/inequality) between two points in the coordinate system. While this
would be ideal from a cryptographic point of view, it would not be useful for spatial databases, since no ``efficient'' indexing would be possible on objects.

\subsection{System Model}

We consider the following system: we assume the existence of mobile nodes $\cal M$ (e.g., mobile devices, sensors) which ``publish'' periodic location reports to a Geofencing service that is hosted in the cloud and consists of multiple database servers.
In the sequel, we denote by ${\cal D}_i$ the $i$-th database server; for clarity of presentation, we also denote by $\cal D$ the ``logical'' database comprising the various database servers. For simplicity and without loss of generality, we assume a 2D bounded area where the nodes can freely move. Each node ${M}_{i} \in {\cal M}$ periodically sends location beacons to the database servers. These beacons consist of tuples of the form $\langle \mathrm{ID}_i, {\mathrm{Loc}}_i\rangle$, where $\mathrm{ID}_{i}$ is an identifier of ${M}_{i}$,
${\mathrm{Loc}}_i$ is a vector of the last movement performed by ${M}_{i}$ (e.g., $\mathrm{Loc}$ could comprise the last and the current coordinates of ${M}_{i}$). We assume that the movement of node ${M}_{i}$ is characterized
by a velocity distribution ${\cal V}_{i}$, and a path distribution ${\cal P}_{i}$. For simplicity and without loss of generality, we assume that the movement of ${M}_{i}$ between two reported coordinates
corresponds to a straight line. Current Geofencing applications require indeed that nodes report their last and the current coordinates; by doing so, the Geofencing server can
be a stateless server which does not memorize
the last coordinate of each node. As we show in this paper, this also serves to increase the privacy of the entire system.

We assume that $\cal D$ offers location-based services to customers, denoted in the sequel by $\cal S$. Here, we assume that customers can ``subscribe'' to events that
occur within a specific sub-area of the map for instance, customers want to be notified when users enter/exit their subscribed Geofences.

In the sequel, we denote by ${S}_{i}$ the $i$th subscriber in $\cal S$; upon receiving a location report from node ${M}_{i}$, $\cal D$ checks if
${M}_{i}$'s reported coordinates are located within a subscribed area. More specifically, $\cal D$ relies on existing spatial database functionality which can efficiently compute the intersection between a line and polygons.
If the movement vector of a node results in non-empty intersection, $\cal D$ issues a notification message to ${S}_{k}$ (e.g., using a URL of ${S}_{k}$ stored at $\cal D$).

\subsection{Adversarial Model and Security Requirements}\label{subsec:attack}

Throughout our analysis, we assume that the nodes are trusted to report their locations correctly. That is, we assume that these devices cannot be compromised by the adversary.
Moreover, we assume that the cloud providers (operating the database servers) are honest-but-curious. More specifically, we assume that each database server will correctly follow the protocol (i.e., authenticate the nodes, output correct notifications) but is
interested in acquiring information about the locations of the nodes in the system, and about the queries that are issued by the customers. Ideally, different database servers do not collude; this assumption especially holds when the database servers
are hosted by different clouds (e.g., Amazon, Google). Moreover, we assume that the adversary cannot physically track the
mobile users to acquire information about their movements.
Finally, we assume that the adversary is computationally bounded (i.e., she cannot acquire secrets, break secure encryption functions, etc.).

As mentioned earlier, the main premise behind our work is to design a privacy-preserving solution for a Geofencing service hosted in the cloud, \emph{without} incurring any modifications on the database servers, and while
ensuring that $\cal D$ does not learn any meaningful information about the location of the users and subscriptions in the system. Since the adversary can compromise a database server, we can express
these security properties using the following requirements:

\vspace{1 em} \noindent \textbf{Requirement 1---Confidentiality of Stored Records:} Each database server should not learn any meaningful information about the stored subscriptions.
This can be ensured if the transcript of interaction between $\cal D$, $\cal M$, and $\cal S$ is (computationally) independent of the actual subscription coordinates.

\vspace{1 em} \noindent \textbf{Requirement 2---Confidentiality of Queries:} Similar to Requirement (1), each data\-base server should not learn any meaningful information about the location of the nodes.
This includes the direction of the movement, the trajectory taken by each node, the distance travelled by each node, etc.

Recall that both Requirements (1) and (2) should be achieved without compromising the functionality of the database server, i.e., while enabling efficient geo-spatial indexing, search over encrypted data, etc. (cf. Section~\ref{subsec:geo}).

\vspace{1 em}  Note that both Requirements (1) and (2) can only ensure confidentiality of the input/output, but do not prevent the possible correlation between
the inputs and outputs of the database when subject to location queries by the nodes. This is the case since $\cal D$ can learn whether a given publish event matches an encrypted subscribe event.
In this work, we do not aim at preventing such information leakage. As far as we are aware, the literature features a number of solutions for this problem. These include delaying some queries to ensure $k$-anonymity~\cite{Guha:2012:KLP:2228298.2228317,Sweeney:2002:AKA:774544.774553,Sweeney:2002:KAM:774544.774552}, relying on
bogus queries/subscription to probabilistically provide input/output unlinkability~\cite{Kido05ananonymous}, among many others.

\section{PrivLoc: Privacy-preserving Outsourced Geofencing Services}\label{sec:solution}

In this section, we introduce \sol, our solution which enables the privacy-preserving outsourcing of Geofencing services to the cloud and we thoroughly analyze its security and privacy provisions.

\subsection{Overview of PrivLoc}\label{subsec:overview}

\sol{} requires that $\cal D$ only implements the standard Geofencing functionality specified in Section~\ref{subsec:geo} without any modification, given inputs from the mobile users. Thus, we see $\cal D$ as a Geofencing service hosted in the cloud. Nevertheless, \sol{} ensures that no meaningful information about the location of the devices and subscriptions in the system is leaked to any entity, including $\cal D$.

In order to achieve these goals, \sol{} relies on a trusted server $\cal T$, which mediates the exchange of information between the devices/subscribers on one side, and $\cal D$ on the other side. More specifically,
$\cal T$ translates both the mobile device locations, and the subscriptions into ``scrambled'' inputs that are then stored and processed by $\cal D$.
Although the inputs are subsequently hidden from $\cal D$, \sol{} ensures that they can be processed using existing geo-spatial indexing algorithms and always result in a correct database lookup. More specifically, the various operations undergone in \sol{} are:
\begin{itemize}
\item Upon receiving each subscription request, $\cal T$ translates the subscription requested by the subscribers into the appropriate coordinates and stores them at ${\cal D}$.
\item Upon receiving each sensor location beacon, $\cal T$ translates the location into the appropriate coordinates and only forwards the transformed beacon to ${\cal D}$.
\end{itemize}

We see $\cal T$ as an additional service which is run locally to prevent information leakage towards spatial database servers hosted in clouds, such as Google, Amazon, etc.
Clearly, for our solution to be effective, the overhead on $\cal T$ should be minimal. Indeed, in \sol, $\cal T$ simply
has to apply a series of transformations on the received subscriptions and location reports. We stress at this point that the role of $\cal T$ can be emulated by the mobile nodes and the subscribers themselves, in case
these entities pre-share keys. Otherwise, the presence of a trusted service which orchestrates the key management among users and subscribers of the database is required (see~\cite{Chow:2007:EPC:1784462.1784477,Bamba:2008:SAL:1367497.1367531,Gruteser:2003:AUL:1066116.1189037,Kalnis:2007:PLI:1313047.1313203,DBLP:conf/infocom/WangXHZLX12} for similar assumptions).

\sol{} introduces
two granularity levels when encrypting locations. On the coarse-granular level, \sol{} relies on a pseudo-random permutation to emulate a strong encryption function, while on the fine-granular level \sol{} relies on
a weaker notion---order-preserving encryption---to preserve the relative location of closely-related points within a sub-area---referred to as \emph{tiles}
---so that the spatial operations performed by $\cal D$ can be applied within each tile
without modification. As we describe in Section~\ref{subsec:spec}, \sol{} relies on multiple database servers to ensure that any two consecutive location reports can be fitted to at least one tile. The combination
of our techniques ensure
that location reports pertaining to each tile can be processed using existing functionality of $\cal D$, without revealing the location or direction of movements. In the following paragraphs, we discuss in details the various translation operations and the load incurred by $\cal T$ in \sol.

\subsection{PrivLoc: Protocol Specification}\label{subsec:spec}

\sol{} unfolds as follows. $\cal T$ first proceeds to dividing the original map by a regular grid into fixed-size square tiles. For privacy reasons, the area covered by a tile should fall above a given threshold. In the sequel, we assume the global system parameters shown in Table~\ref{tab:parameters}. Furthermore, \sol{} makes use of the following functions:

\begin{table}[tbp]
\centering
\begin{tabular}{l | p{0.8\textwidth}}
\hline
$\SECPAR$ & the security parameter, e.g. 128 bit\\
$\LENGTH$ & the width and height of a tile\\
$\COLUMNS$& the number of tiles in a column, i.e. the height of the map\\
$\ROWS$ & the number of tiles in a row, i.e. the width of the map\\
\hline
\end{tabular}
\caption{Global system parameters of \sol.}
\label{tab:parameters}
\vspace{-1.5 em}
\end{table}

\begin{itemize}
\item a pseudorandom function $\mathrm{PRF}: {\cal K} \times \{0,1\}^* \times \ell \rightarrow \{0,1\}^\ell$ , which can be realized by a HMAC construction.
\item a pseudorandom permutation $\mathrm{PRP}: {\cal K} \times \{0,1\}^N \rightarrow \{0,1\}^N$,\\where $N=\ceil{\log_2(\ROWS \COLUMNS)}$. Exemplary PRP constructions can be found in~\cite{black2002ciphers}.
\end{itemize}

In the setup phase, $\cal T$ generates key $\MK$ as a uniformly random bit string of a length depending on the intended security level $\SECPAR$. 

As mentioned earlier, \sol{} requires that $\cal T$ encrypts the coordinates reported by devices before transmitting them to $\cal D$. Besides hiding location information, one important goal here is to
hide the fine-grained user movements (i.e., direction and distance of movement).
Recall that, in our setting, the subscribers are interested in knowing whether a device has crossed a Geofence. For the database servers to determine that, it is therefore
necessary for them to be able to compare the relative coordinates between the origin and the destination of every sensor movement segment along with the boundaries of Geofences (or subscriptions).

\sol{} achieves the aforementioned goals by dividing the map into tiles and distorting the original coordinate system into three different variant maps (for the reasoning why, see following paragraphs).
By doing so, \sol{} encrypts the location of each tile within the maps, but ensures that every received location report can be fitted to at least one full tile in one of the maps. Subscriptions that cross more than one tile
are also subsequently split to several smaller Geofences that are completely contained in a single tile. Within that tile, \sol{} further distorts
the direction and distance of the user movement while enabling the spatial database to find all intersections between the movement vector and translated Geofences in the distorted tile using existing indexing techniques.

More specifically, \sol{} applies the {\sf encrypt} procedure (Algorithm~\ref{alg:encrypt}); {\sf encrypt} takes as input real world coordinates $(x,y)$ and transforms them into obfuscated coordinates $(\newx, \newy)$.
{\sf encrypt} is executed on \emph{(i)} movement vectors to translate the movement end-points, and on \emph{(ii)} subscriptions to translate the south-west, and the north-east coordinates that define the minimum bounding box of each Geofence.

The algorithm {\sf encrypt} consists of four main routines:
\begin{itemize}
\item {\sf permuteTiles} is used to divide the map into equal-sized tiles and permute these tiles. By doing so, {\sf permuteTiles} hides the location of the devices within the map.
\item {\sf rotateTile} and {\sf flipTile} are used to rotate and flip each tile in the distorted map. Both routines serve to hide the direction of devices' movements within the original map.
\item {\sf OPE} is used to hide the distance between any two locations within each tile.
\end{itemize}

Here, {\sf permuteTiles}, {\sf rotateTile}, and {\sf flipTile} hide the location and direction of the movement of each device from $\cal D$, while {\sf OPE} distorts the distances within each tile.
All four routines, however, enable $\cal D$ to rely on existing indexing techniques (c.f. Section~\ref{subsec:geo}) to compare different locations \emph{within each tile}. 

\begin{algorithm}[t]
\caption{Coordinate Encryption in \sol}
\label{alg:encrypt}
\begin{algorithmic}[1]
  \Require{Coordinates $x,y$, map offset $z$, master key $\MK$}
  \Ensure{Encrypted location $\newx, \newy$}
  \Procedure{encrypt}{$x,y,z,\MK$}
	\State $(\mathit{Num}, dx, dy) \gets \Call{coordinatesOnTile}{x,y,z}$
  \State $\mathit{Num} \gets \mathit{PRP}({\MK};\ROWS\times\COLUMNS;\mathit{Num})$
	\State $(dx, dy) \gets \Call{rotateTile}{dx, dy, \mathit{Num},\MK}$
	\State $(dx, dy) \gets \Call {flipTile}{dx, dy, \mathit{Num},\MK}$
	\State $(dx, dy) \gets \Call {OPE}{dx, dy,Num,\MK}$
	\State $\newx \gets~(\mathit{Num}~\mathbf{ mod }\, \ROWS) \times \LENGTH + dx$
	\State $\newy \gets~(\mathit{Num}~\mathbf{ div }\, \ROWS) \times \LENGTH + dy$
	\State \textbf{return} $\newx, \newy$
    \EndProcedure
\end{algorithmic}
\end{algorithm}

\paragraph{\underline{Hiding Movements within a Tile}\\}

Recall that \sol{} operates on movement vectors reported by the users.
Upon receiving the vector from devices, $\cal T$ encrypts the vector, by applying the
{\sf encrypt} procedure in Algorithm~\ref{alg:encrypt}, to both the start and end-point of the movement vector. These points are represented by 2D coordinates $(x,y)$. The encryption function
invokes the procedure {\sf coordinatesOnTile}, which returns the tile number $\mathit{Num}$ of the tile where the point $(x,y)$ lies on, and the relative
coordinates $(dx, dy)$ on the respective tile. As the map is split in $\ROWS \times \COLUMNS$ tiles, we have $\mathit{Num}\in\mathbf{Z}_{\ROWS\times\COLUMNS}$. This coordinate translation is achieved by means of Algorithm~\ref{alg:divide}.

As the first step of hiding the location of these points within the map, $\cal T$ permutes the tiles using the key $\MK$ by applying the pseudorandom function: $\mathit{PRP}({\MK};\ROWS\times\COLUMNS;\mathit{Num}).$ Note that $\MK$ is
only held by the trusted server $\cal T$. This function computes a pseudorandom permutation of the numbers in $\mathbf{Z}_{\ROWS\times\COLUMNS}$ and outputs the new position $\mathit{Num}$ for the tile $T$.

\begin{algorithm}[t]
\caption{Coordinate Translation}
\label{alg:divide}	
\begin{algorithmic}[1]
\Require{Coordinates $x,y$ and offset $z$ of the respective map}
\Ensure{Num of the tile where $x,y$ lies and relative coordinates $dx, dy$ on the tile}
\Procedure{coordinatesOnTile}{$x,y$}
\State $x0 \gets (x - z~\mathbf{mod}~\LENGTH \times \COLUMNS)~\mathbf{div}~\LENGTH$
\State $y0 \gets (y - z~\mathbf{mod}~\LENGTH \times \ROWS)~\mathbf{div}~\LENGTH$
\State $\mathit{Num}\gets (x0 + y0 \cdot \ROWS)$
\State $(dx, dy) \gets x~\mathbf{mod}~\LENGTH, y~\mathbf{mod}~\LENGTH$
\State \textbf{return}$(x0,y0,dx,dy)$
\EndProcedure
\end{algorithmic}
\end{algorithm}

To further hide the linkability between two consecutive location reports (i.e., to prevent leakage of movement information), $\cal T$ first rotates each tile using {\sf rotateTile} (rotation
chosen at random among 0, $\pi/2$, $\pi$, $3\pi/4$), and then (individually) flips them using {\sf flipTile}, in order to obfuscate the
direction of the movement. Both operations result in a transformation entropy of three bits per tile. When combined with {\sf permuteTile}, this results in a total of $3\floor{\log_2(nm)}$ bits of entropy per tile.

{\sf rotateTile} is described in Appendix~\ref{alg:rotate}. {\sf rotateTile} takes the output of {\sf permuteTile}; the position of the tile $T$ is not changed in this algorithm. The
rotation angle of $T$ is determined using the key-based pseudo-random function $\text{PRF}$.

$\cal T$ then applies tile flipping (Algorithm~\ref{alg:flip}). Given the outputs of {\sf rotateTile}, {\sf flipTile} outputs the newly flipped coordinates $(\newx, \newy)$.
Note that {\sf flipTile} is analogous to {\sf rotateTile}, except that the coordinates are transformed by a mirror matrix.

In order to hide the distances of movements executed by devices, \sol{} further relies on the use of order-preserving encryption (OPE) within each tile.
Order-preserving encryption has the property that the relative order among coordinates is preserved after encryption. That is, let $\mathrm{OPE}(p)$ denote the order-preserving encryption of plaintext $p$.
Then, the following holds:
\vspace{-1 em}
\begin{equation*}
\left\{
\begin{array}{rl}
  \mathrm{OPE}(p_1) \leq \mathrm{OPE}(p_2) & if p_1 \leq p_2, \\
\mathrm{OPE}(p_1) > \mathrm{OPE}(p_2) & if p_1 > p_2.
\end{array}
\right.
\end{equation*}

\begin{algorithm}[tbp]
\caption{Tile Flipping}
\label{alg:flip}
\begin{algorithmic}[1]
\Require{Coordinates $dx,dy$ and $\mathit{Num}$}
\Ensure{Coordinates $dx, dy$ after flipping the tile}
\Procedure{flipTile}{$x,y,\MK$}
\State $\mathit{flip} \gets$ PRF($\MK$; ``flip'', $\mathit{Num}$; $1$)
\If $\mathit{flip} = `1'$
	\State $dx' \gets (\LENGTH-1) - dx$
\EndIf
\State \textbf{return}$(dx', dy)$
\EndProcedure
\end{algorithmic}
\end{algorithm}

\vspace{-1 em}\paragraph{\underline{Hiding Cross-Tile Movements}\\}

Clearly, {\sf encrypt} can only hide the trajectory, and distance exhibited by location reports from $\cal D$ when they fall within the same tile. However, if two consecutive
location reports cross the boundary of a single tile and are matched to different tiles, then \emph{(i)} $\cal D$ cannot compare the relative location advertised by these reports
and \emph{(ii)} $\cal D$ might be able to guess that these tiles correspond to physically connected tiles in the original map.

\begin{figure}[tb]
	\centering
		\includegraphics[width=0.6\linewidth]{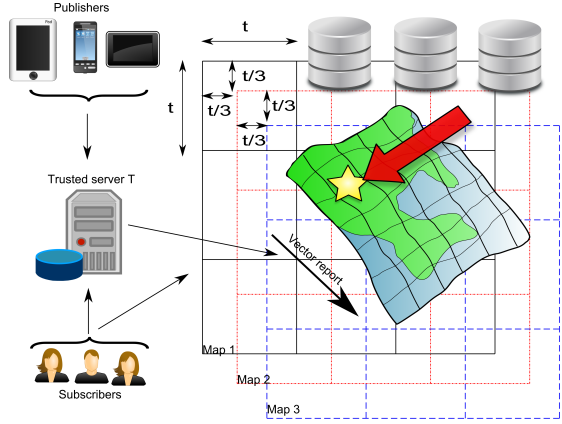}
		\caption{Example of three tilings of the original map. Notice here that the movement vector crosses the boundaries of tiles in ``Map 1'' (shown in solid lines) and ``Map 2'' (shown in dotted lines) but can be fitted complexity within a tile in ``Map 3'' (shown in dashed lines).}
	\label{fig:tiles}
\end{figure}

Thus, a query for a movement which crosses the tile boundary must be avoided in \sol. To achieve that, \sol{} relies on more than one tiling of the same map (see Figure~\ref{fig:tiles}); although a movement
might cross the boundary of one map tiling, \sol{} ensures that there is at least one tiling, in which the movement falls completely within a single tile. Here, by different map tilings, we refer to slightly shifted variants of the original map, which
have been processed (i.e., using the {\sf encrypt} routine) by means of different keys (cf. Figure~\ref{fig:tiles}). As shown in Figure~\ref{fig:tiles}, the maximum
magnitude of a movement vector that \sol{} accepts is bounded by $l=t/3+\epsilon$, where $t$ is the length of one tile. Thus, all vectors with magnitude $d(x,y)<r=t/3§$ can be located within one complete tile stored at least on
one of the three servers. Within that tile, this enables the comparison between all closely located points $x$ and $y$, where $d(x,y)<t/3$, while ensuring that
the movement from location $x$ to $y$ does not cross the tile\footnote{Thus, this ensures that an adversary who can observe consecutive movements cannot acquire information about the relative position of the actual tiles in each transformed map.}.
An example of a map with three tilings is depicted in Figure~\ref{fig:tiles}. As shown in Figure~\ref{fig:tiles}, we point out that relying on two tilings of the same map is not enough. This is the case since there exist
movement vectors which can still cross two different tiles in any two map tilings. However, it is easy to show that given three shifted tilings of the same map, there is no point where all boundaries intersect, thus ensuring that every movement
can be fitted to one tile pertaining to at least one tiling. This is exactly why \sol{} requires the presence of three database servers ${\cal D}_1$, ${\cal D}_2$, and ${\cal D}_3$, which store subscriptions pertaining to the three map tilings.

All received subscriptions are
stored encrypted with three different keys $k_1$, $k_2$ and $k_3$ on all three servers respectively.  Whenever $\cal T$ receives a location report, it will choose to query the database server ${\cal D}_i, i \in \{1,2,3\}$ for whom
this report falls in a full tile (see Appendix~\ref{alg:chooseserver} for the algorithm).
Here, $\cal T$ can efficiently find out which server to query for each movement (by essentially performing two integer divisions, cf. Section~\ref{sec:implementation}).

As a by-product, we point out that the reliance on multiple database servers
in our scheme inherently achieves load balancing of the load on the servers and increases the load capacity of the entire system.
This is the case since $\cal T$ only queries one server for each received location report.

\paragraph{\underline{Encrypting Subscriptions}\\}

We now proceed to describing how $\cal T$ encrypts subscriptions and stores them at the database servers $\cal D$. Clearly, for $\cal D$ to be able to use existing functionality to compute the intersections of the movements with subscriptions, the subscriptions must be encrypted using the same routine that is used to process the location reports.

\begin{figure}[tb]
	\centering
		\includegraphics[width=\linewidth]{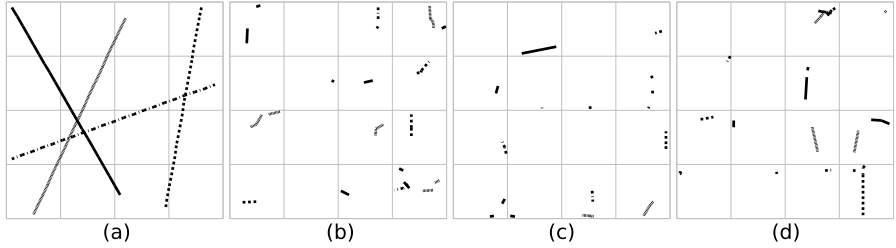}
		\caption{Exemplary run of {\sol}~with four different movements. Each path displayed in (a) consists of 20 movement reports, each of a length of 30\% of the tile width. We depict the snapshots seen by database servers ${\cal D}_1$, ${\cal D}_2$, and ${\cal D}_3$ in (b), (c) and (d), respectively. Note that the marking of the vectors in (b)-(d) is only for visualisation, since different location reports cannot be linked by the database. }
	\label{fig:example}
\end{figure}

More specifically, $\cal T$ uses the {\sf encrypt} routine to encrypt the north-east and the south-west coordinates which define the minimum bounding box for each subscription.
This is done using the three keys $k_1$, $k_2$, and $k_3$, respectively. The resulting encrypted coordinates are stored in database servers ${\cal D}_1$, ${\cal D}_2$, and ${\cal D}_3$ (see Figure~\ref{fig:example}).
Here, $\cal T$ must ensure that:
\begin{itemize}
	\item In case a subscription crosses the boundaries of tiles for the map tiling at a database server, the subscription needs to be split into parts, that each completely fit within one tile of the corresponding map tiling. This process
has to be repeated for each of the three servers independently. While this incurs additional storage overhead per server to store the tiles, we show in Section~\ref{sec:implementation} that the storage blowup incurred by \sol{}
can be, to a large extent, tolerated in realistic settings.
	\item $\cal T$ batches the upload of $\bar{k}$ subscriptions to the three database servers. Here, $\bar{k}$ denotes a desired privacy threshold. As we show later, this ensures that database servers cannot temporally correlate various subscriptions.
\end{itemize}

\subsection{Security Analysis}\label{subsec:def}

Before analyzing the security of \sol, we capture Requirements (1) and (2) (cf. Section~\ref{subsec:attack}) using the following security game, $\mathrm{Priv}^{\cal A}$, which involves a p.p.t adversary ${\cal A}$ and a
challenger $\cal C$. In our game, $\cal C$ simulates location reports and subscription requests and emulates the role of the trusted server ${\cal T}$ which interacts with $\cal D$. More specifically, the $\mathrm{Priv}^{{\cal A}}$ game unfolds as follows.
\begin{description}
\item [Setup.] The challenger ${\cal C}$ sets up 3 database servers ${\cal D}_1$, ${\cal D}_2$ and ${\cal D}_3$ and generates the respective master keys $k_1$, $k_2$ and $k_3$ using the $\mathrm{setup}$ routine.
\item [Run.] The challenger ${\cal C}$ simulates location reports and subscriptions. More specifically, $\cal C$ simulates the presence of $N$ nodes with distribution ($\cal V$, $\cal P$), and $S$ subscriptions.
On input a location report $(x,y)$, $\cal C$ executes {\sf chooseServer} and {\sf encrypt}$(x,y,k)$.
\item [Compromise.] ${\cal A}$ chooses \emph{one} server ${\cal D}^{{\cal A}}_i$, $i \in \{1,2,3\}$ at time $t$.
Starting from time $t$, ${\cal A}$ acquires a trace consisting of all inputs and outputs to/from ${\cal D}^{{\cal A}}_i$.
\item [Challenge.] ${\cal A}$ then chooses a location report $r$ that arrived at $\tilde{t} > t$ and sends $r$ as a challenge to $\cal C$.
\item [Response.] Upon reception of $r$, $\cal C$ locates the node $M_k$ which issued $r$, and randomly flips a bit $b$. If $b=0$, then $\cal C$ sends to ${\cal A}$ a trajectory of movement vectors followed by node $M_k$.
Otherwise, if $b=1$, $\cal C$ creates a randomly generated trajectory comprising of movement vectors derived from the distributions ${\cal V}_k$ and ${\cal P}_k$.
\end{description}

We define the advantage of ${\cal A}$ in the above game $\mathrm{Priv}^{{\cal A}}$ by:
\begin{equation*}
\mathrm{Adv}_{\mathrm{Priv}^{{\cal A}}}=\mathrm{Prob}[b' \leftarrow {{\cal A}} : b=b']
\end{equation*}

\begin{definition}\label{def1}
\textbf{Location Tracking}\\
We say that a system prevents $\epsilon$-tracking if $\mathrm{Adv}_{\mathrm{Priv}^{{\cal A}}} = \frac{1}{2} + \epsilon.$
\end{definition}

Clearly, a system perfectly prevents location tracking when $\epsilon$ is negligible. Definition~\ref{def1} captures Requirements (1) and (2) in Section~\ref{subsec:attack}. Here, ${\cal A}$ can compromise
one database server. After observing the
transcript of interactions with the compromised server, our goal is to prevent ${\cal A}$ from acquiring information about the trajectory followed by any node $M_k$ of her choice.
This is captured by the fact that the probability that ${\cal A}$ can distinguish the trajectory adopted by $M_k$ from any random path is negligible. Notice that $\mathrm{Priv}^{\cal A}$ captures
the ability of the adversary to infer information about the trajectory taken by users using side-information like the velocity of movement, the distance traveled, etc.

\vspace{1 em} \noindent \textbf{Security Analysis: }We informally analyze the security of our scheme with respect to the aforementioned $\mathrm{Priv}^{\cal A}$ game.

Notably, our goal is to show that an adversary $\cal A$ which compromises one database server $\overline{{\cal D}_i}$ and has access to the inputs/outputs of $\overline{{\cal D}_i}$ cannot acquire meaningful information
about the trajectory of users. Recall that the communication between users, subscribers, and $\cal T$ is performed over confidential and authenticated channels, which does not
give $\cal A$ any advantage in acquiring information about the (plaintext) location reports and the subscriptions in the system.

In analyzing the advantage of $\cal A$ in $\mathrm{Priv}^{\cal A}$, two cases emerge:
\begin{description}
\item[\emph{1) Analyzing single location reports: }] Having received an encrypted location vector from $\cal T$, we note that $\cal A$ cannot infer the location of the corresponding user whom generated it in the original map. This is the case
since the keyed PRP used in the {\sf permuteTiles} function ensures that the adversary cannot guess the actual location of the tile which hosts the report in the original map. Moreover, OPE ensures that $\cal A$ cannot
acquire the actual distance travelled by the user while the keyed tile rotation and flipping ensure that $\cal A$ cannot guess the direction of the user movement given the received location vector.
\item[\emph{2) Correlating two or more location reports: }] Recall that $\overline{{\cal D}_i}$ will only receive location reports which correspond to a complete tile given the tiling hosted by $\overline{{\cal D}_i}$. As mentioned earlier, the combined use of the permuted tile location of the report, the OPE-obfuscated distance
travelled by the user, and tilted/rotated movement direction does not offer any distinguisher for $\cal A$ to correlate two or more location reports.
\end{description}

Similarly, it is easy to show that $\cal A$ cannot acquire any meaningful information about the subscriptions in the system. Note that when tiling and permuting tiles at each server, each subscription
might be split into a number of smaller subscriptions within each tile. However, assuming that the system hosts a number of subscriptions, this does not give any advantage to $\cal A$ in inferring information about the subscriptions.
Note, here, that $\cal A$ can acquire considerable information as $\cal T$ populates $\overline{{\cal D}_i}$. For instance, it is straightforward for $\cal A$ to guess with high probability that consecutive subscriptions
in $\overline{{\cal D}_i}$
correspond to the same actual subscription which was subsequently split by $\cal T$ to prevent subscriptions from crossing tile borders. This is exactly why \sol{} requires that $\cal T$ batches the processing of $\bar{k}$ (genuine) subscriptions
at a time. On one hand, this enhances the anonymity of the subscriptions, and on the other hand, this prevents $\cal A$ from correlating stored encrypted subscriptions (e.g., map them to the same subscription) and acquiring information about the tile permutation.

As mentioned in Section~\ref{subsec:attack}, \sol{} can only ensure confidentiality of the input/output, and does not prevent the correlation among
the inputs and outputs of $\overline{{\cal D}_i}$. That is, $\cal A$ can learn whether a given location report has triggered a response from $\overline{{\cal D}_i}$ and therefore corresponds
to a user entering/exiting a Geofence. The literature features a number of solutions for this problem (see~\cite{Guha:2012:KLP:2228298.2228317,Sweeney:2002:AKA:774544.774553,Sweeney:2002:KAM:774544.774552,Kido05ananonymous}).

\section{Implementation \& Evaluation}\label{sec:implementation}

In this section, we implement \sol{} and we evaluate the performance of our implementation in comparison to the setting where the Geofencing service is not outsourced to the cloud, but is locally offered using an existing spatial database.
\vspace{-0.5 em}
\subsection{Implementation Setup}

We implement the \sol{} service using Java. In our implementation, we set $\SECPAR=128$, we adapt the OPE from~\cite{Boldyreva:2009:OSE:1533674.1533691}, and
we use HMAC-SHA1 as the PRF in the {\sf rotateTile} and {\sf flipTile} routines. The tile permutation is achieved
by an array permutation given a secure random number generator. 
In our setup, we deploy \sol{} on an 8-core Intel Xeon E3-1230 with 16 GB RAM; the various clients issuing location reports and/or subscriptions were co-located with $\cal T$ on the same machine.

We chose a square $100km \times 100km$ map in which the maximum size of a subscription is $100m \times 100m$ (this was also the tile size in
our implementation). Conforming with Section~\ref{subsec:spec}, we set the maximum
 movement distance of a user between two consecutive location reports to $\frac{t}{3}\approx 30m$. We rely on the random waypoint model to simulate node mobility. More specifically, for each user in the system, we assume a random initial location
 in the map; for each subsequent movement, the movement vector angle is chosen randomly from [0,360] degrees, while the distance of each movement is chosen randomly from [0,30] meters. Throughout our evaluation, we assume a realistic setting where
 the ratio of location reports to incoming subscription requests is 19:1.\footnote{We conducted experiments where the ration between where location reports and subscriptions is 1:1. Our results were similar to the setting featuring a ratio of 19:1.}

Based on this setup, we measure the throughput achieved by \sol{} with respect to the average latency incurred on $\cal T$ due to a location report or a subscription. To increase the load on $\cal T$, we vary the number of concurrent clients from 1 to a maximum of 256. In our experiments, we are interested in assessing the performance of
\sol{} when the links between $\cal T$ and the database servers are not the bottleneck. For that purpose, we short-circuit the database servers and we abstract away the time to upload the location reports by $\cal T$ to the database servers.
Under these settings, we compare the throughput and latency achieved by \sol{} to the traditional case where the spatial database is entirely hosted locally; here, we compare the performance
of \sol{} to \emph{(i)} a stand-alone MySQL spatial database (with R-tree indexing) and \emph{(ii)} a PostgreSQL database (with GiST indexing). Both databases were deployed within our 8-core Intel Xeon E3-1230 with 16 GB RAM; we report their performance with different subscription table sizes (i.e., initial number of records). We also evaluated \sol{}
when compared to MongoDB. Since the performance exhibited MongoDB was far inferior to that of MySQL and PostgreSQL, we omit these measurements from our evaluation.

\subsection{Evaluation Results}

\begin{wrapfigure}{r}{0.5\textwidth}
\vspace{-5 em}
\includegraphics[width=0.5\textwidth]{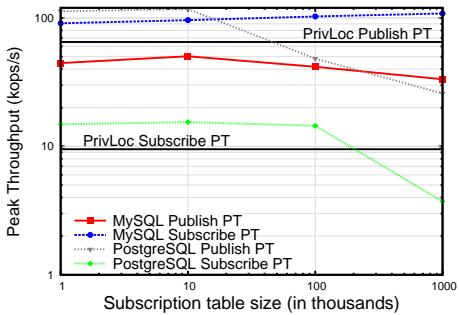}
\vspace{-2 em}
\caption{Comparison between \sol{} and local MySQL and PostgreSQL databases.}
\label{fig:blow-up}
\vspace{-2 em}
\end{wrapfigure}

We start by evaluating the performance of \sol{} w.r.t. to the MySQL and PostgreSQL databases while varying the initial number of stored subscriptions. This experiment captures the performance of \sol{} compared to locally hosted spatial databases as the number of subscription populating these databases increases with time. For that purpose, we measure the peak throughout (PT) exhibited by \sol{} and compare it with that of MySQL, and PostgreSQL respectively, in settings where the initial number of stored subscriptions varies from 1,000 to 1,000,000. Our results are depicted in Figure~\ref{fig:blow-up}. Our findings show that the PT of \sol{} is superior to MySQL when processing location reports, irrespective of the number of stored subscriptions. Here, \sol{} outperforms PostgreSQL as the number of subscriptions stored by the PostgreSQL database increases beyond 50,000 records. This is also the case when processing subscription requests. Nevertheless, our results indicate that, even when the initial number of subscriptions stored in the MySQL, and PostgreSQL databases is as low as 1,000, the relative PT achieved by \sol{} can be easily tolerated.

\begin{figure}[tbp]
\hfill
\subfigure[Average latency for processing a location report by $\cal T$ with respect to the throughput.]{\includegraphics[width=0.47\linewidth]{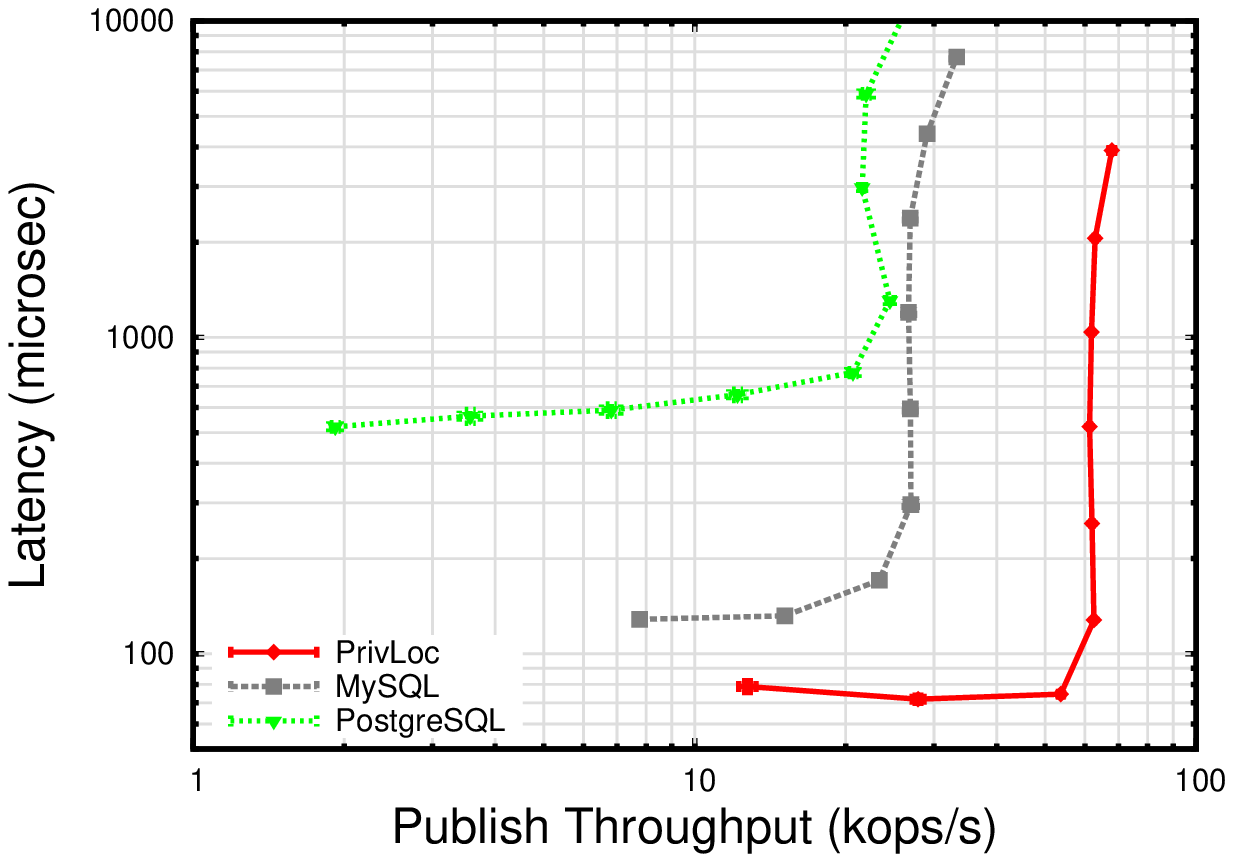}
\label{fig:eval-publish}}
\hfill
\subfigure[Average latency for processing a subscription by $\cal T$ with respect to the throughput.]{\includegraphics[width=0.47\linewidth]{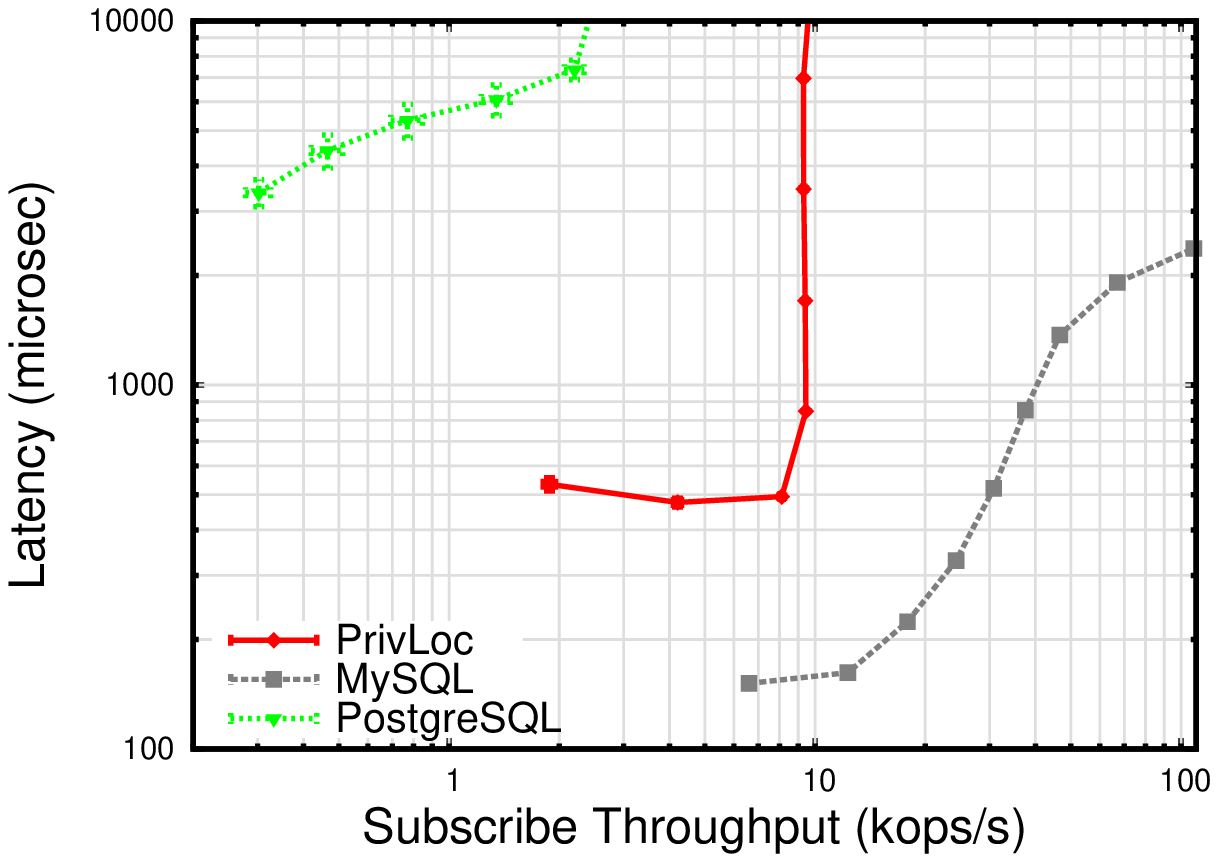}
\label{fig:eval-subscribe}}
\hfill
\caption{Throughput and latency in \sol. Here, we assume a subscription table size of 1,000,0000 records. Each data point in our measurements is averaged over 10 independent measurements; we present the corresponding 95\% confidence intervals.}
\label{fig:throughput}
\vspace{-1 em}
\end{figure}

In a second experiment, we evaluate the relative performance of \sol{} in the realistic case where
the Geofencing service has been running for some time, and has accumulated the subscriptions from a large number of subscribers. To simulate this case, we insert 1,000,000 subscription records in the databases.
Figure~\ref{fig:throughput} depicts the latency incurred in \sol{} for the processing of location reports and subscriptions in the system with respect to the
achieved throughput (measured in the number of 1000 operations
per second). For comparison purposes, we also include the performance achieved by a local MySQL and a PostgreSQL spatial database in the same setup.
Our results (cf. Figure~\ref{fig:eval-publish}) show that \sol{} is at least twice faster than locally processing location reports on both spatial databases. Moreover, the peak throughput achieved in \sol{} is more than 2 times higher than that
achieved by the MySQL and PostgreSQL databases. However, our results in Figure~\ref{fig:eval-subscribe} show that the latency and peak throughput achieved by \sol{} are modest when inserting subscriptions, compared to the local MySQL database.
This is due to the blow-up in the number of subscriptions. Notably, since $\cal T$ splits the map into small tiles, a subscribed area might
be further split by $\cal T$ if it crosses multiple tiles. Note that each subscribed area results in an average of 6.75 subscribed areas which will be encrypted
by \sol{} and pushed to the $n=3$ database servers. This blow-up is dependent on the number of database servers, and the size of the
subscribed area with respect to the size of the tiles. More specifically, the average area of a subscription in our setup is $1/4$ of the area of a tile; this means that the
average number of partitions of a subscribed area on each database server is $N_{p} = 1\cdot\frac{1}{4} + 2\cdot\frac{2}{4}  + 4\cdot\frac{1}{4}= 2.25$. The total blowup in terms of subscriptions is subsequently $3N_{p}= 6.75$.

In spite of the storage blowup, our findings nevertheless show that \sol{} considerably outperforms a local PostgreSQL database.

\vspace{-1 em}
\section{Related Work}\label{sec:related}

In what follows, we briefly overview existing contributions in the area.
In~\cite{Barkhuus03location-basedservices}, Barkkuus and Dey show that users were concerned about the ability of services to track them.

Most privacy-enhancing solutions for location-based services rely on a trusted ``location anonymizer'' service which hides the location of users. These services either
provide $k$-anonymity~\cite{Guha:2012:KLP:2228298.2228317,Gedik:2008:PLP:1340084.1340202,Sweeney:2002:AKA:774544.774553,Sweeney:2002:KAM:774544.774552,Samarati:2001:PRI:627337.628183} or spatial/temporal cloaking with an area of
interest~\cite{Chow:2007:EPC:1784462.1784477,Bamba:2008:SAL:1367497.1367531,Gruteser:2003:AUL:1066116.1189037,Kalnis:2007:PLI:1313047.1313203,DBLP:conf/infocom/WangXHZLX12}. A number of solutions rely on inserting fake queries in order to prevent a database server from learning the actual location reports (e.g.,~\cite{Kido05ananonymous}). While these
solutions provide $k$-anonymity, they incur significant additional costs on the database server. Other solutions rely on location perturbation/obfuscation; these solutions map the location reports to a set of pre-defined landmarks~\cite{Hong:2004:APU:990064.990087} or blur
the user location into a spatial area using linear transformations~\cite{Bamba:2008:SAL:1367497.1367531,Gruteser:2003:AUL:1066116.1189037,Kalnis:2007:PLI:1313047.1313203,Mokbel:2006:NCQ:1182635.1164193,Yiu:2010:ESS:1825238.1825264}
Such solutions indeed hide the location of users but might affect the accuracy of the location-based service. Moreover, these solutions
can only hide the location of a user, but do not aim at hiding the user movement.

To prevent location tracking, Gruteser and Liu~\cite{Gruteser:2004:PPC:1435702.1437404} propose disclosure control algorithms which hide users' positions in sensitive areas and withhold path information that
indicates which areas they have visited. Other schemes rely on Private Information Retrieval (PIR) algorithms in order to enable privacy-preserving queries in spatial databases~\cite{Ghinita:2008:PQL:1376616.1376631,Olumofin:2010:AEQ:1881151.1881157}. PIR schemes allow a querier to retrieve information
from a database server without revealing what is actually being retrieved
from the server. However, these solutions
are computationally intensive and require modifications to the database server in order to process the blurred location queries.

In~\cite{terminology}, Pfitzmann \emph{et al.} define unlinkability and privacy in pseudonymous systems.
Dwork~\cite{diffprivacy} define differential privacy and quantify information leakage
from the query access of individuals. In~\cite{locprivacy}, Shokri \emph{et al.} quantify location privacy using the error of the adversarial estimate from the ground truth. In~\cite{Diaz02,Serj02}, various entropy-based metrics are introduced to assess the communication privacy in anonymous
networks.

\vspace{-1 em}
\section{Conclusion}\label{sec:conclusion}

In this paper, we proposed a novel solution, \sol, which enables privacy-preserving outsourcing of location-based services to the cloud without
 leaking any meaningful information to the cloud provider. \sol{} goes one step beyond existing solutions in the area and targets Geofencing services where
 users send a vector of their movements for the service provider to detect whether a user has crossed a given Geofence.
 We analyze the security and provisions of \sol{} and show that \sol{} does not leak information about the location, movement, trajectory, and/or velocity of the users to the Geofencing database. Our
 evaluation of \sol{} shows that the overhead incurred by our solution can be largely tolerated in realistic deployment settings.

\vspace{-1 em}
\section*{Acknowledgements}
\vspace{-0.5 em}
This work was supported by the EU FP7 SMARTIE project (contract no. 609062).
\vspace{-1 em}

\bibliographystyle{plain}
\bibliography{publish_subscribe}
\vspace{-0.5 em}
\appendix
\footnotesize
\vspace{-1 em}
\section{Tile Rotation in \sol}\label{alg:rotate}
\vspace{-0.5 em}

\begin{algorithmic}[1]
\Require{Coordinates $dx,dy$ and $\mathit{Num}$}
\Ensure{Coordinates $dx, dy$ after rotating the tile}
\Procedure{rotateTile}{$x,y,\MK$}
\State rotation $\gets$ PRF($\MK$; ``rotate'', $\mathit{Num}$; $2$)
\If {rotation $=$ `00'}
	\State $dx' \gets dx$
	\State $dy' \gets dy$
\ElsIf {$rotation = `01'$}
	\State $dx' \gets dy$
	\State $dy' \gets (\LENGTH-1)-dx$
\ElsIf {$rotation = `10'$}
  \State $dx' \gets (\LENGTH-1)-dx$
	\State $dy' \gets (\LENGTH-1)-dy$
\ElsIf {$rotation = `11'$}
  \State $dx' \gets (\LENGTH-1)-dy$
	\State $dy' \gets dx$
\EndIf
\State \textbf{return} $(dx', dy')$
\EndProcedure
\end{algorithmic}

\vspace{-1.5 em}
\section{Querying the appropriate database server in \sol}\label{alg:chooseserver}
\vspace{-0.5 em}
	
\begin{algorithmic}[1]
\Require{Coordinates $x0,y0$ and $x1,y1$ describing a movement}
\Ensure{Server where $x0,y0$ and $x1,y1$ are on the same tile or $\bot$}
\Procedure{{\sf chooseServer}}{$x,y$}
\ForAll{${\cal D}_i,z_i$ with $i = 1,2,3$} \Comment{Server {\cal D} with tile offset $z$}
\State $nx \gets (x0 - z_i~\mathbf{mod}~\LENGTH \times \COLUMNS)~\mathbf{div}~\LENGTH$
\State $ny \gets (y0 - z_i~\mathbf{mod}~\LENGTH \times \ROWS)~\mathbf{div}~\LENGTH$
\State $\mathit{Num0}\gets (nx + ny \cdot \ROWS)$
\State $nx \gets (x1 - z_i~\mathbf{mod}~\LENGTH \times \COLUMNS)~\mathbf{div}~\LENGTH$
\State $ny \gets (y1 - z_i~\mathbf{mod}~\LENGTH \times \ROWS)~\mathbf{div}~\LENGTH$
\State $\mathit{Num1}\gets (nx + ny \cdot \ROWS)$
\If{Num0 = Num1} \State \Return $i$ \EndIf
\EndFor
\State \Return $\bot$
\EndProcedure
\end{algorithmic}

\end{document}